\RequirePackage{fix-cm}

\documentclass[onecolumn]{svjour3}
\usepackage{amsmath}
\usepackage{graphicx}
\usepackage{epstopdf, mathtools, mathdesign}
\usepackage{float, color}
\usepackage{latexsym}
\usepackage{amssymb}
\usepackage{array}
\usepackage{xcolor, color}

\title{R\'enyi Holographic Dark Energy Models In Multidimensional Universe}

\author{A. Saha        \and
        A. Chanda \and
        S. Dey \and 
        B. C. Paul \and
        S. Ghose }
\institute{A. Saha \at
              Physics Dept., Jalpaiguri Govt. Engg. College, Japlaiguri 735102 \\
              \email{arindamjal@gmail.com}          
           \and
           A. Chanda  \at  Physics Dept., University of North Bengal, Rajarammohunpur 734013\\
           \email{anirbanchanda93@gmail.com}           \and
           S. Dey \at Physics Dept., University of North Bengal, Rajarammohunpur 734013\\
           \email{sagardey231@gmail.com}           \and
           B. C. Paul \at Physics Dept., University of North Bengal, Rajarammohunpur 734013\\
           \email{bcpaul@nbu.ac.in}           \and
           S. Ghose \at ICARD, Physics Dept., University of North Bengal, Rajarammohunpur 734013\\
           \email{dr.souvikghose@gmail.com}           
}

\vspace{3in}
\begin{document}

\maketitle

\begin{abstract}
R\'enyi Holographic Dark Energy (RHDE) Models has been studied in different dimensions considering a flat FRW metric, subsequently exploring the nature of evolution of different cosmological parameters. It is seen that the number of dimensions affect the evolution of the cosmological parameters and the transition of the universe from decelerating to accelerating phase. The statefinder diagnosis has been performed and it is seen that the model corresponds to the $\Lambda CDM$ model for the present and future universe.
\keywords{R\'enyi Holographic Dark Energy \and Higher Dimensions \and Stability}
\end{abstract}

\vspace{5in}

\pagebreak 
\section {Introduction:}
\label{sec:intro}

It has been two decades since supernovae observation suggested that  the present universe is passing through a phase of accelerating expansion. Subsequent observations only strengthened this belief \cite{riess,perlm,perlm2,perlm3}. The present phase of acceleration can be successfully modeled by modifying the Einstein-Hilbert action in the first place \cite{nojde1,cappode}. Alternatively, some dynamical form of energy with negative pressure, termed as dark energy, is thought to dominate the present epoch, leading to the accelerated rate of expansion \cite{freese,overduin,fritzsch,basilakos,sola,ratra1988,gibb,armendariz2001,wetterich2017,escamilla2013}. Dynamical dark energy or modification of gravity is not demanded by the late phase of acceleration as it can be explained with a simple cosmological constant($\Lambda$) also \cite{carroll2001cosmological}. The motivation for considering other approaches lies in the fact that the tiny value of the $\Lambda$ as suggested by observation can not be explained from any fundamental theory and there is a huge mismatch of about $~10^{60}$. 
Another alternative approach, vastly studied in literature, is to explore dimensions more than four \cite{chodos,shafi,wetterich,ACCETTA,kitaura,bcp}. Higher-dimensional models are considered to serve the dual purpose of solving cosmological riddles such as late-time acceleration and at the same time providing us with a unified model of the fundamental forces of nature. The idea behind these models is that, although undetectable at the present epoch, the extra dimensions play a crucial role in driving the inflation in the early universe or the acceleration at a later time. One of the earliest examples of higher-dimensional models is the five-dimensional Kaluza-Klein theories \cite{kaluza,klein,leekk,appelkk}. Since then a plethora of higher dimensional theories like the string theory \cite{cline2006}, superstring theory \cite{lidsey1999}, braneworld models \cite{brax2004} and others have been explored. Although none of these models stands out as satisfactory, several of them show great prospect both in the cosmological and the fundamental physics concepts. In the present work we compare general FRW universes in different dimensions without diving into any specific higher dimensional model.

Recently, the holographic principle has emerged as an essential proposal of quantum gravity \cite{cohen,susskind}. The principle states that, \emph{"the number of degrees of freedom of any physical system should scale with the boundary and not with the volume"}. The holographic principle leads to the Holographic dark energy (HDE) models in the cosmological context with several interesting features. Cosmology itself does not offer any unique choice of the long-distance cutoff, also known as the infrared cut off (IR cut off) for any of the HDE models \cite{granda2008}. One of the most natural choices of IR cut off is the Hubble horizon (for cosmology in generaliged HDE and with arbitrarty cut off see \cite{nojiriarb,nojirisym}). However, in the original model of HDE (OHDE) if one considers the Hubble radius as IR cut off and the Bekenstein entropy, then both dark matter (DM) and OHDE are scaled with the same function of the scale factor. Hsu subsequently noted that HDE cannot explain the present accelerated phase of expansion \cite{hsu2004}. Later Zimdahl and Pavon showed that an interaction between the dark sectors (dark energy and dark matter) would lead to an accelerating universe in the late time \cite{zimdahl2007}. {\it Very recently Nojiri \emph{et. al.} \cite{nojiriunify} have advanced an interesting idea of unifying an early inflationary epoch and the late time accelerating phase of the universe using the holographic approach.} branches of HDE emerged as other non-extensive entropy conditions were explored (see \cite{miao2011} for a detailed review). Recent works suggest one such model, the R\'enyi Holographic Dark Energy (RHDE hereafter), yields accelerated expansion in Kaluza-Klein framework even without any interaction \cite{SAHA}. \\
The motivation of the paper is to compare non-interacting R\'enyi Holographic dark energy (RHDE) in different higher dimensional contexts. {\it But, presently, no specific higher dimensional model is dealt with. As suggested by numerous observations, the universe is assumed to be completely homogenous and isotropic in large scale. Evolution of different cosmological parameters are studied and compared for different dimensions. A four dimensional universe is emphasized in a sense that any higher dimensional theory must admit an effective four dimensional scenario where diagnostics are physically meaningful. However, it is worth exploring in this context if the presence of extra dimension would alter the evolution of different parameters. As the effect of dark energy is more prevalent in late time, the present work is more focused on the late time evolution of the universe.}  \\

The paper is organised as follows: In sec. (~\ref{sec:rhde}), a general idea of R\'enyi entropy and R\'enyi dark energy density in higher dimensions is presented. Sec. (~\ref{sec:fileq}) advances the relevant field Equation in higher dimensions. The cosmological evolution is discussed in the sec. (~\ref{sec:evuniv}). In sec. (~\ref{sec:stabdia}), stability and diagnostics of the model are studied, and finally, the findings are discussed in the sec. (~\ref{sec:disc}).

\section{General idea of R\'enyi entropy and R\'enyi dark energy density in higher dimension:}
\label{sec:rhde}
It has been argued that the systems including long-range interactions are in accord with the generalized entropy formalisms based on the power law distribution of probabilities. Recently, such entropies have been employed to build new holographic dark energy models that can describe the cosmic evolution from matter dominates era to the current accelerating universe.\\

In non-extensive thermodynamics the Tsallis entropy ($S_T$) \cite{tsalis,biro2013q,czinner2016renyi,belin2013holographic} for a set of $W$ states is defined  as 
\begin{equation}
\label{tsaen}
S_{T}=k_{B} \frac{1- \sum_{i=1}^{W}p_{i}^{q}}{q-1} \; \; {\left(\sum_{i=1}^{W}p_{i}=1; \; q\in \rm I\!R \right)},
\end{equation}
where $p_i$ is the probability associated to the $i^{th}$ microstate with $\sum_{i}p_{i}=1$ and $q$ is any real number. Tsallis holographic dark energy has been extensively studied in the context of cosmology \cite{sharma2020diagnosing,varshney2019statefinder,thdemain} The original R\' enyi entropy ($S_{R}^{Org}$), on the other hand, is defined as \cite{tsalis}:
\begin{equation}
\label{rsaen}
S_{R}^{Org}=k_{B} \frac{ln\sum_{i=1}^{W}p_{i}^{q}}{q-1}=\frac{1}{1-q}ln\left[1+(1-q)S_{T}\right].
\end{equation}
It is interesting that both the eqs.(\ref{tsaen}) and (\ref{rsaen}) lead to Boltzmann-Gibbs entropy for $q=1$. Recently, it has been proposed that the Benkenstein-Hawking ($S_{BH}$) entropy too is a kind of non-extensive entropy which leads to a novel type of R\' enyi entropy \cite{czinner2016renyi} which is given by:
\begin{equation}
\label{rsaen1}
S_{R}=\frac{1}{\delta}ln\left(1+\delta S_{BH} \right),
\end{equation}
where $\delta=1-q$ and for $\delta=0$, $S_{R}=S_{BH}$.
In cosmology, R\'enyi holographic dark energy is extensively considered\cite{sharma2020statefinder,dubey2020diagnosing}. It is shown in \cite{moradpour2017accelerated} that when R\'enyi entropy is employed at the horizon it results in an accelerating universe in the Rastall framework. In the present work, the R\' enyi entropy is considered for describing the entropy on the Hubble horizon. As $S_{BH}=\frac{A}{4}$ (where $A$ is the area of the horizon), one obtains
\begin{equation}
\label{rsaen2}
S_{R}=\frac{1}{\delta}ln\left(1+ \frac{A}{4} \; \delta \right).
\end{equation}
In cosmological setup the relation $dE_T = TdS$ can be considered as the first law of thermodynamics for the cosmological horizon \cite{cai}. Here, $E_T$ denotes the total energy of the universe, $T= \frac{1}{2\pi L}$,  the Cai-Kim temperature for a system with IR cutoff $L$, corresponds to the temperature of the de-Sitter horizon and apparent horizon in flat FRW cosmology, $S$ is the horizon entropy. Also in DE dominant present universe, one can use the assumption $dE_D=\rho_D dV\propto dE_T = TdS$, where $E_D$ denotes the energy content for the DE representative.\\
Note that the apparent horizon is a proper casual boundary for universe in agreement with the thermodynamical laws and thus the energy-momentum conservation law. Hence considering the Hubble radius as the IR cutoff ($L=H^{-1}$), the energy density for DE candidate is obtained as:

\begin{equation}
\label{eden}
\rho_D = \frac{C'H}{2\pi}\frac{dV}{dS},  
\end{equation}
where $C'$ is the proportionality constant.

Again  $n$ dimensional volume $(V_n)$ and surface area $(S_{n-1})$ of radius $R$ is given by

\begin{equation}
\label{hdvol}
V_n=\frac{\pi ^{\frac{n}{2}}}{\Gamma\left(\frac{n}{2}+1\right)}R^{n},  
\end{equation}

and

\begin{equation}
\label{hdarea}
S_{n-1}=\frac{2\pi ^{\frac{n}{2}}}{\Gamma\left(\frac{n}{2}\right)}R^{n-1},  
\end{equation}

Using the equation (\ref{eden}), (\ref{hdvol}), and (\ref{hdarea}) the energy density for RHDE can be expressed as: 

\begin{equation}
\label{energyden}
\rho_D= \frac{B^2H^2}{1+\frac{\delta K}{H^{(n-1)}}},
\end{equation}

where $B^2=\frac{C^2}{8\pi}$, \;\;\;\; $C^2= \frac{C'(n-1)K'K_{1}^{\frac{1}{n-1}}}{n}$,\\

$K'=\frac{2\pi^\frac{n}{2}}{\Gamma\left(\frac{n}{2}\right)}\left[\frac{\Gamma\left(\frac{n}{2}+1\right)}{\pi ^{\frac{n}{2}}}\right]^{\frac{n-1}{n}}$, $K_1=\frac{\Gamma\left(\frac{n}{2}+1\right)}{\pi ^{\frac{n}{2}}}$\\
and $K=\frac{K'}{4K_{1}^{\frac{n}{n-1}}}$.

\section{Field Equation for cosmology in Higher Dimension}
\label{sec:fileq}
The FRW metric in $(1+1+d)$ dimension is given by

\begin{equation}
\label{hdmetric}
dS^2=dt^2-a^2(t)\left[\frac{dr^2}{1-kr^2}+r^2(dX_d)^2\right],
\end{equation}
where $a(t)$ denotes the scale factor and $k=0,\pm1$ represents the curvature parameter for flat and closed (open) universe.\\
We consider higher dimensional flat universe filled by two types of cosmic fluids. The total energy density is then $\rho=\rho_D +\rho_m$, where $\rho_D$ corresponds to dark energy and $\rho_m$ is for matter including cold dark matter with $\omega_m=0$.\\ The Einstein field equations for flat FRW universe in $(1+1+d)$ dimension are as follows:

\begin{equation}
\label{hdfdequn1}
H^2=\frac{16\pi}{d(d+1)}(\rho_D +\rho_m)
\end{equation}
and
\begin{equation}
\label{hdfdequn2}
H^2+\frac{2\dot{H}}{d+1}=-\frac{16\pi}{d(d+1)}P_d
\end{equation}
Here, we also assume that there is no energy exchange between the cosmic fluids. For non-interacting fluids, the energy conservation equations are as follows:
\begin{equation}
\label{hdcontm}
\dot{\rho}_{m}+H(d+1)\rho_{m}=0,
\end{equation}
\begin{equation}
\label{hdcontl}
\dot{\rho}_{D}+H(d+1)\rho_{D}(1+\omega_{D})=0.
\end{equation}
The dimensionless density parameters are defined as: 
\begin{equation}
\label{dpdef}
\Omega_m=\frac{\rho_m}{\rho_{cr}}, \; \;  \Omega_{D}=\frac{\rho_D}{\rho_{cr}},
\end{equation}
where $\rho_{cr}=\frac{d(d+1)H^2}{16\pi}$. Eq. (\ref{hdfdequn1}) can be written in terms of density parameters as:
\begin{equation}
\label{denequ}
\Omega_m +\Omega_{D}=1.
\end{equation}
The ratio of the energy densities is given by:
\begin{equation}
\label{endenra}
r'=\frac{\rho_m}{\rho_D}=\frac{\Omega_m}{\Omega_D}=\frac{1-\Omega_D}{\Omega_D}.
\end{equation}

\subsection{Generalized Holographic Dark Energy}
In the holographic principle, the holographic dark energy density is proportional to the inverse squared infrared (IR) cut off $L_{IR}$, which is related to the holographic DE density as,
\begin{equation}
    \rho_{HDE}=\frac{3C^{2}}{\kappa^{2}L_{IR}^{2}},
\end{equation}
where $\kappa^{2}=8\pi G$ and $C$ is a numerical constant that acts as a free parameter. The IR cut off is assumed to be the particle horizon $L_{P}$ or the future event horizon $L_{F}$, which are defined respectively as,
\begin{equation}
    L_{P}\equiv a\int_{0}^{t}\frac{dt}{a},\;\;\;\;\;\;\;\;\;\; L_{F}\equiv a\int_{t}^{\infty}\frac{dt}{a}.
\end{equation}
The Hubble parameter can now be expressed in terms of $L_{P}$, $L_{F}$ and their time derivatives. Taking time derivative of the above equation we get,
\begin{equation}
    H(L_{P},\dot{L_{P}})=\frac{\dot{L_{P}}}{L_{P}}-\frac{1}{L_{P}}, \;\;\;\;\;\;\;\; H(L_{F},\dot{L_{F}})=\frac{\dot{L_{F}}}{L_{F}}+\frac{1}{L_{F}}.
\end{equation}
The general form of the cut off is given by (Ref.\cite{nojiriuorg}),
\begin{equation}
    L_{IR}=L_{IR}(L_{P},\dot{L_{P}},\ddot{L_{P}},......,L_{F},\dot{L_{F}},\ddot{L_{F}},.....a).
\end{equation}
The IR cutoff depends on other factors as well such as the Hubble parameter, the Ricci Scalar and their derivatives. However, they can be transformed to either $L_{P}$ and its derivatives or $L_{F}$ and its derivatives. The above-mentioned cutoff can be chosen to be equivalent to a general covariant gravity model as
\begin{equation}
    S=\int d^{4}\sqrt{-g}F(R,R_{\mu\nu}R^{\mu\nu},R_{\mu\nu\rho\sigma}R^{\mu\nu\rho\sigma},\Box R,\Box^{-1}R,\nabla_{\mu}R\nabla^{\mu}R,.......).
\end{equation}
These expressions can be used to derive a generalized IR cutoff for the RHDE.\\
For a higher dimensional spacetime one can write equation (17) as,
\begin{equation}
    \rho_{HDE}=\frac{d(d+1)C^{2}}{2\kappa^{2}L_{IR}^{2}}
    \end{equation}
where $d=2$, gives the usual $4$ dimensional results. If one uses the Bekenstein-Hawking (BH) entropy and the first law of thermodynamics ($TdS=dQ$, where $S$ is the entropy and $Q$ is the increase of heat in the region inside the cosmological horizon) then the Friedmann equation for a $(1+1+d)$ dimensional spacetime can be obtained. However, instead of the BH entropy one may consider the R\'enyi entropy (eqn. 4) and apply the first law of thermodynamics to obtain the following relation,
\begin{equation}
    \frac{\dot{\rho}}{H(d+1)}=\frac{d}{8\pi}\frac{1}{(1+\frac{\delta}{4}k_{A}H^{-d})}\dot{H}
\end{equation}
where, $k_{A}=\frac{2\pi ^{\frac{n}{2}}}{\Gamma\left(\frac{n}{2}\right)}$, with $n=(d+1)$ and we have considered the radius of the cosmological horizon to be equal to the inverse of the Hubble rate ($R=\frac{1}{H}$).\\
The integration of the above equation yields an expression for the DE density as,
\begin{equation}
    \rho=\frac{d(d+1)}{8\pi}\int{\frac{HdH}{(1+\frac{\delta}{4}k_{A}H^{-d})}}+C_{0}
\end{equation}
where $C_{0}$ is an integration constant. One can compare equations (22) and (24) and argue that RHDE belongs to the generalized HDE family even in higher dimensions. The corresponding infrared cutoff $L_{R}$ in this case depends on the dimension and is given by,
\begin{equation}
    \frac{d(d+1)C^{2}}{2\kappa^{2}L_{R}^{2}}=\frac{d(d+1)}{8\pi}\int{\frac{HdH}{(1+\frac{\delta}{4}k_{A}H^{-d})}}+C_{0}.
\end{equation}
The above equation is too complicated to solve for a general $d$ value. However, solutions for a particular dimension and the corresponding value for the generalized cutoff can be computed. The EoS parameter can be obtained from the conservation of equation of the HDE density $\rho_{HDE}$ :
\begin{equation}
    \Omega_{HDE}^{R}=-1-\frac{\rho_{D}}{(d+1)H\rho_{D}}=-1+\frac{2}{(d+1)HL_{R}}\frac{dL_{R}}{dt},
\end{equation}
where $L_{R}$ is given by eqn. (25). Thus we conclude that in higher dimensions also, $\Omega_{HDE}^{R}$ is equivalent to $w_{D}$ as obtained in \cite{nojirisym} for four dimensions.

\section{Evolution of Universe With RHDE}
\label{sec:evuniv}

Taking the time derivative of eq. (\ref{hdcontm}) and eq. (\ref{hdcontl}) we can reach
\begin{equation}
\label{hubdot}
\frac{\dot{H}}{H^2}=-\left(\frac{d+1}{2}\right)\left(1+\omega_D \Omega_D \right)
\end{equation}

In $(1+1+d)$ dimension one can easily obtain the expression of density parameter as:

\begin{equation}
\label{hddenp}
\Omega_D=\frac{2C^2}{d(d+1)\left(1+\frac{\delta K}{H^d}\right)},
\end{equation}

Now, taking the time derivative of eq. (\ref{hddenp}) we have:

\begin{equation}
\label{hdrhdebddot}
\frac{\dot{\rho_D}}{\rho_D}=\frac{\dot{H}}{H}\Big[\frac{2C^{2}(d+2)-d^{2}(d+1)\Omega_{D}}{2C^{2}}\Big].
\end{equation}

Combined with relation $\Omega_{D}^{'}=\frac{\dot{\Omega_{D}}}{H}$ eq. (\ref{hdrhdebddot}) gives the evolution equation of RHDE density parameter as:

\begin{equation}
\label{denprime}
\Omega_{D}^{'}=-\frac{d(d+1)\Omega_D}{4C^2}\left[2C^2-d(d+1)\Omega_D\right]\left(1+\omega_D\omega_D\right),
\end{equation}

where $(')$ denotes derivative with respect to $\ln a$ i.e, $\Omega_{D}^{'}=\frac{d\Omega_D}{d\ln a}$.

\begin{figure}
    \centering
    \includegraphics[width=1.0\textwidth]{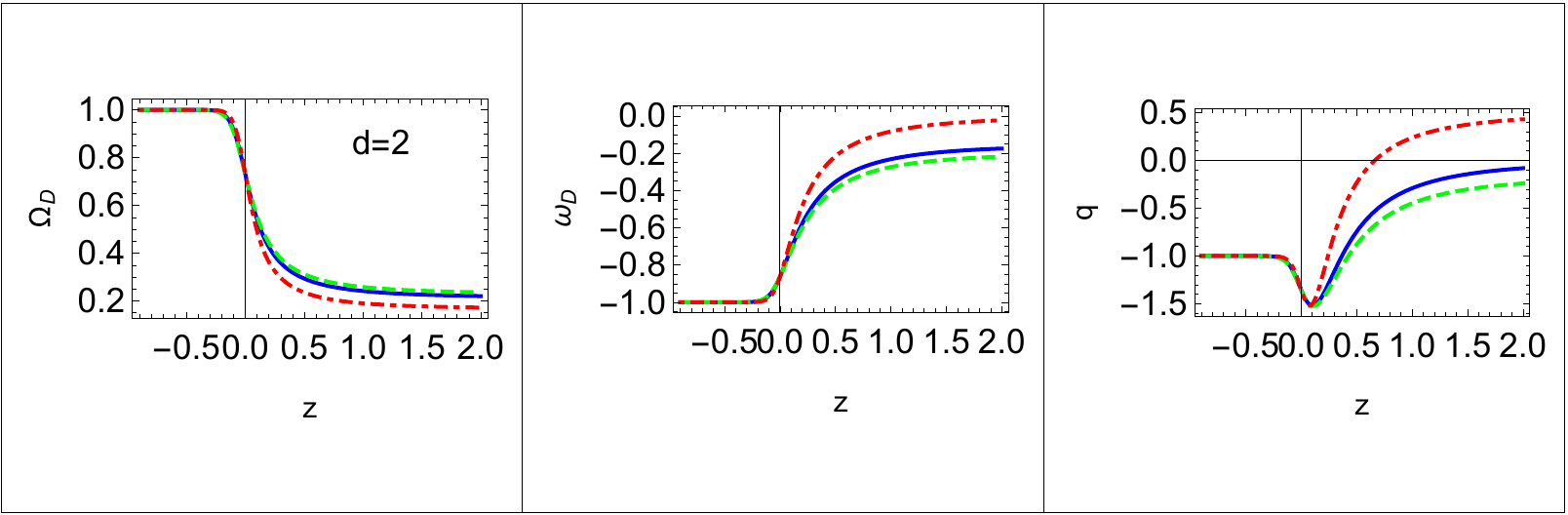}
    \caption{Evolution of cosmological parameters for different $\delta$ values in four dimensions with $\delta=-0.0001$ (Green, Dashed), $\delta=-0.0002$ (Blue, Solid) and $\delta=-0.0005$ (Red, Dot-dashed).}
    \label{fig:param1}
\end{figure}

EoS parameter is obtained from eqs. (\ref{hdcontl}), (\ref{hubdot}), and (\ref{hdrhdebddot}): 
\begin{equation}
\label{eoshd}
\omega_D= \frac{d \Big[2 C^{2}-d (d+1) \Omega_{D}\Big]}{-2 C^{2} (d+2) \Omega_{D}+4 C^{2}+(d+1) d^2 .\Omega_{D}}
\end{equation}

Using eqs. (\ref{hubdot}) and (\ref{eoshd}) the deceleration parameter can be written as: 
\begin{equation}
\label{decparahd}
q=\Big(\frac{d-1}{2}\Big)+\Big(\frac{d+1}{2}\Big)\Big[\frac{d(2C^{2}-d(d+1)\Omega_{D}}{4C^{2}-2C^{2}(d+2)\Omega_{D}+d^{2}(d+1)\Omega_{D}^{2}}\Big]
\end{equation}

\begin{figure}[ht]
    \centering
    \includegraphics[width=1.0\textwidth]{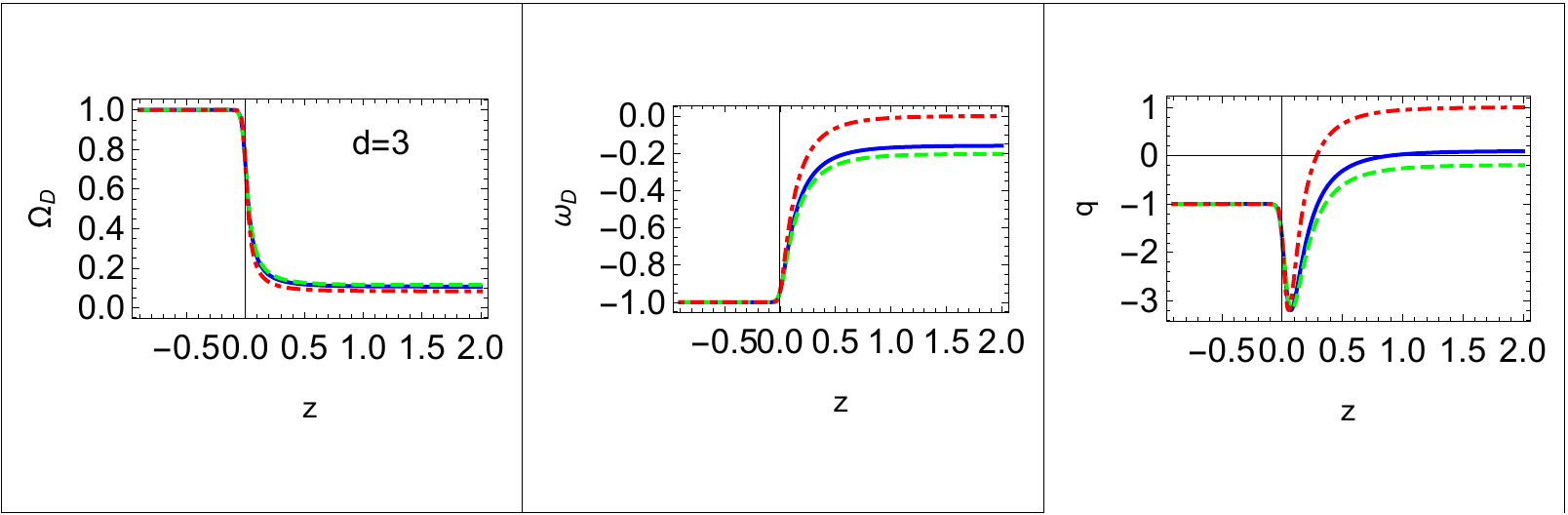}
    \caption{Evolution of cosmological parameters for different $\delta$ values in five dimensions with $\delta=-0.0001$ (Green, Dashed), $\delta=-0.0002$ (Blue, Solid) and $\delta=-0.0005$ (Red, Dot-dashed).}
    \label{fig:param2}
\end{figure}

\begin{figure}[ht]
    \centering
    \includegraphics[width=1.0\textwidth]{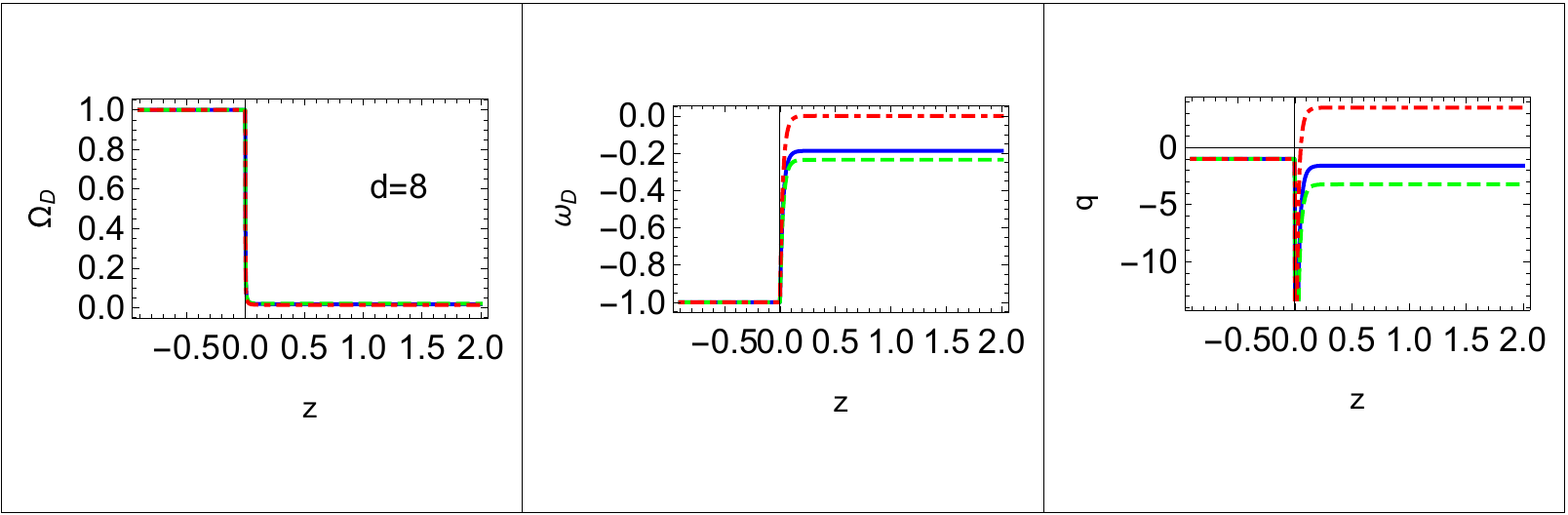}
    \caption{Evolution of cosmological parameters for different $\delta$ values in ten dimensions with $\delta=-0.0001$ (Green, Dashed), $\delta=-0.0002$ (Blue, Solid) and $\delta=-0.0005$ (Red, Dot-dashed).}
    \label{fig:param3}
\end{figure}

In the figs. (\ref{fig:param1})-(\ref{fig:param3}) the cosmological parameters $\Omega_{d}$, $\omega_{d}$ and $q$ have been plotted respectively for different values of $\delta$ and three different dimensions $4$, $5$ and $10$ of the universe, for the initial condition $H(z=0)=0.07 Gy^{-1}$ and $\Omega_{m0}=0.27$. These figures clearly show that, depending on the values of $\delta$ the acceptable behaviour Of $\Omega_{d}$, $\omega_{d}$ and $q$ can be obtained. In addition, comparing figs. (\ref{fig:param1})-(\ref{fig:param3}) with each other, we observe that the changes in the density parameter become sharper as one goes for higher dimension and with the increase in dimension of universe the behaviour of $\Omega_{D}$ with redshift $z$ becomes almost independent of model parameter $\delta$.\\

From the evolution of the EoS parameter $w_{d}$ (figs. (\ref{fig:param1}), fig. (\ref{fig:param2}), and (\ref{fig:param3})), it is evident that the dark energy begins in a quintessence era in early universe and gradually progressing towards the phantom domain $i.e.$ $w<-1$. However in the late universe the EoS parameter attains a value $w=-1$ and never crosses the phantom divide irrespective of the dimensions. These transitions are faster as in an higher dimensional universe.\\

Also, the deceleration parameter $(q)$ starts from positive value at the earlier time and goes to negative in recent past, suggesting a decelerating phase, sutable for structure formation as well as the late-time acceleration of universe. For different value of the model parameter $\delta$, present value of $q$ is seen to vary from $-0.4$ to $-0.1$ which is very much acceptable in light of supernovae data. Moreover, eq. (\ref{decparahd}) suggests that the transition redshift $z_t$ (at which the universe expansion phase is changed from the decelerated matter dominated phase to an accelerated expansion) occurs at the redshift for which $\Omega_D=\sqrt{\frac{2C^2}{d(d+1)}}$. Hence the values of model parameter $\delta$ and diemension of the universe $d$ also affect the value of transition redshift. It is also evident from the figures that the universe remains ever accelerating in the future and this behaviour is independent of both the dimensions and $\delta$


\begin{figure}
\label{fig:param4}
\centering
\includegraphics[width=1.0\textwidth]{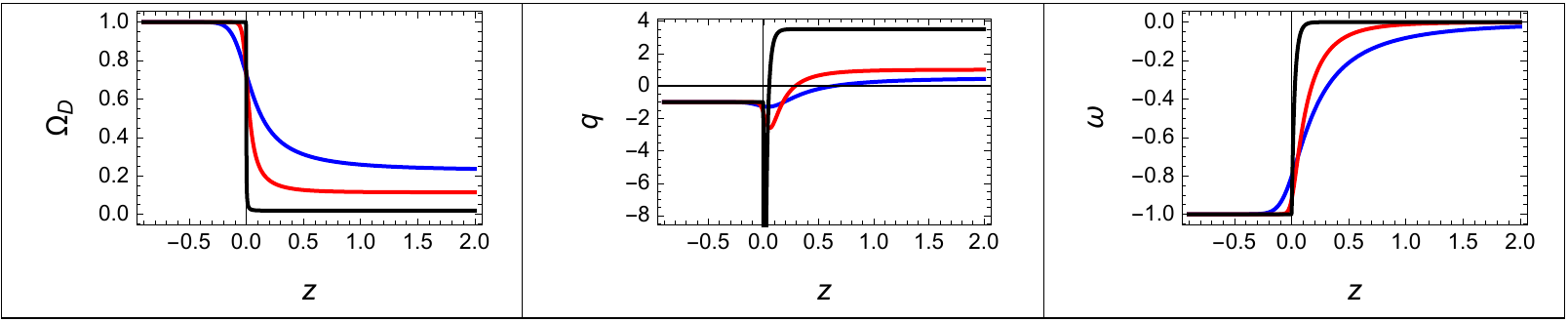}
\caption{Evolution of the density parameter ($\Omega_{D}$), deceleration parameter ($q$) and the EoS parameter ($w$) in higher dimensions. Four dimensional (Blue), five dimensional (Red) and ten dimensional (Black) with $\Omega_{D0}=-0.73$ and $H_{0}=0.07 Gy^{-1}$.}
\end{figure}

\section{Classical Stability and Diagonestics}
\label{sec:stabdia}

\subsection{Classical Stability of RHDE}
\label{sec:clastab}
\begin{figure}[ht]
    \centering
    \includegraphics[width=1.0\textwidth]{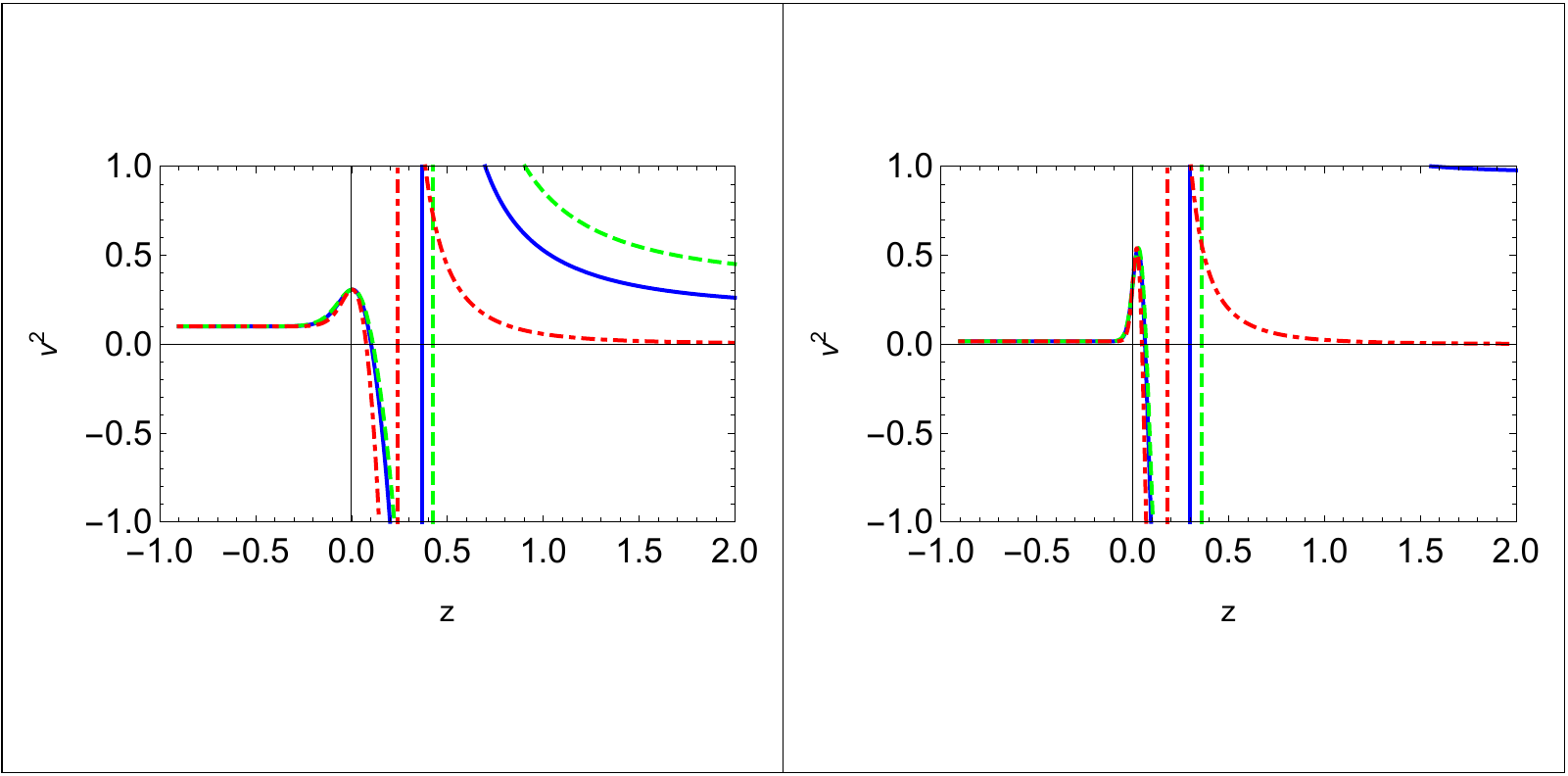}
    \caption{Evolution of the square speed of sound $v^{2}$ with redshift $z$ in four (left) and five (right) dimensions for $\delta=-0.0001$ (Green, Dashed), $\delta=-0.0002$ (Blue, Solid) and $\delta=-0.0005$ (Red, Dot-dashed).}
    \label{fig:sqspeed}
\end{figure}

In this section we study the classical stability of our model against perturbations.  For this the squared speed of sound $v^{2}$ is considered as an indicator of stability. $v^{2}>0$ indicates that the given perturbation propagates in the environment and hence shows the stability a model. The squared speed of sound of in the present model is given by

\begin{equation}
\label{sqss}
\begin{split}
v^2 & =\frac{d\rho_D}{dP_D} \\
    & = \omega_{d} \\
		& + \frac{d^{2}\Omega_{D}(2C^{2}-d(d+1)\Omega_{D})(-4C^{2}(d^{2}-2)}{(2C^{2}(d+2)-d^{2}(d+1)\Omega_{D})(4C^{2}-2C^{2}(d+2)\Omega_{D}+d^{2}(d+1)\Omega_{D}^{2})^{2}} \\ 
		& - \frac{4C^{2}d^{2}(d+1)\Omega_{D}+d^{3}(d+1)^{2}\Omega_{D}^{2}}{(2C^{2}(d+2)-d^{2}(d+1)\Omega_{D})(4C^{2}-2C^{2}(d+2)\Omega_{D}+d^{2}(d+1)\Omega_{D}^{2})^{2}}.
\end{split}
\end{equation}
The above equation indicates that the stability of the model depends on model parameter $\delta$. We have ploted $v^{2}$ against red-shift parameter $z$ for three different dimensions in the fig. (\ref{fig:sqspeed}). It is interesting to note that for some value of model parameter $\delta$ the model shows stability at present time as well as from very recent past and continue to depict the same in future also. We note that as the dimension increases the squared speed of sound shows a discontinuity at a particular redshift however remaining stable in other redshifts depending on $\delta$ values.\\

\subsection{Statefinder Diagnostic}
\label{sec:statf}
Higher dimensions, as they might play crucial role in unifying the natural forces or demystifying cosmic riddles, should be observationally equivalent to a four dimensional universe at present. For example the compact Kaluza-Klein theories achieve this feat by suggesting compactification of the extra dimensions. So, the state finder diagnostics of RHDE is studied for the four dimensional universe. The state finder pair $(r,s)$ is defined as follows \cite{sahni2003statefinder,alam2003exploring}:
\[
r=\frac{\dddot{a}}{aH^3}=\frac{\ddot{H}}{H^3}+\frac{3\dot{H}}{H^2}+1
\]
and
\[
s=\frac{r-1}{3(q-\frac{1}{2})}
\]
Using Eqs. (\ref{hdfdequn1}) and (\ref{hdfdequn2}) the state finder parameters can be expressed as 
\begin{equation}
\label{hdr}
r=1+\frac{(d+1)(d-2)}{2}+\frac{(d+1)}{2}\left[(2d-1)+(d+1)\omega_D\right]\omega_D \Omega_D-\frac{(d+1)}{2}\Omega_D\omega_{D}^{'}
\end{equation}

where,
\begin{equation}
    \label{hdwp}
\omega_{D}'=\frac{\dot{\omega}_{D}}{H}=\frac{d(-4C^{2}(d^{2}-2)-4C^{2}d^{2}(d+1)\Omega_{D}+d^{3}(d+1)^{2}\Omega_{D}^{2})}{(4C^{2}-2C^{2}(d+2)\Omega_{D}+d^{2}(d+1)\Omega_{D}^{2})^{2}}
\end{equation}

\begin{figure}[htbp]
    \centering
    \includegraphics[width=1.0\textwidth]{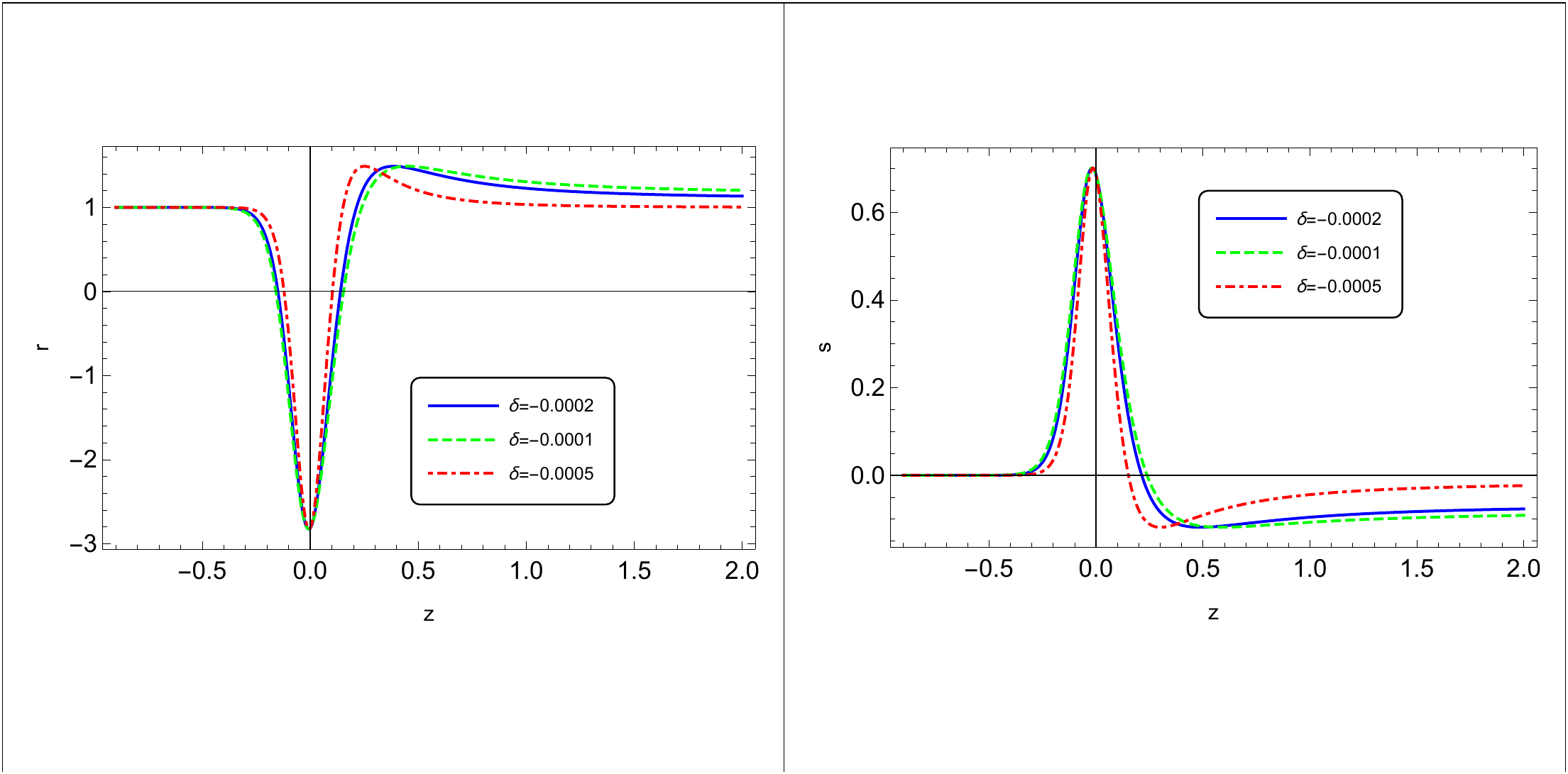}
    \caption{Evolution of the statefinder pair $r$ (left) and $s$ (right) with redshift $z$ in four dimensions for $\delta=-0.0001$ (Green, Dashed), $\delta=-0.0002$ (Blue, Solid) and $\delta=-0.0005$ (Red, Dot-dashed).}
    \label{fig:statef}
\end{figure}

The variation of the statefinder pair $(r,s)$ with redshift $z$ are plotted in case of four dimensions as shown in fig. (\ref{fig:statef}). It is evident from the figure that in four dimension the statefinder pair approach their corresponding $\Lambda CDM$ values in the future. However, the present universe shows significant departure from a  $\Lambda CDM$ model. One interesting thing to note here is that for higher dimensions the statefinder pair show a significant departure from their $\Lambda CDM$ value at a particular positive redshift. However RHDE model corresponds to $\Lambda CDM$ model for $z\le0$.

\begin{figure}%
\includegraphics[width=\columnwidth]{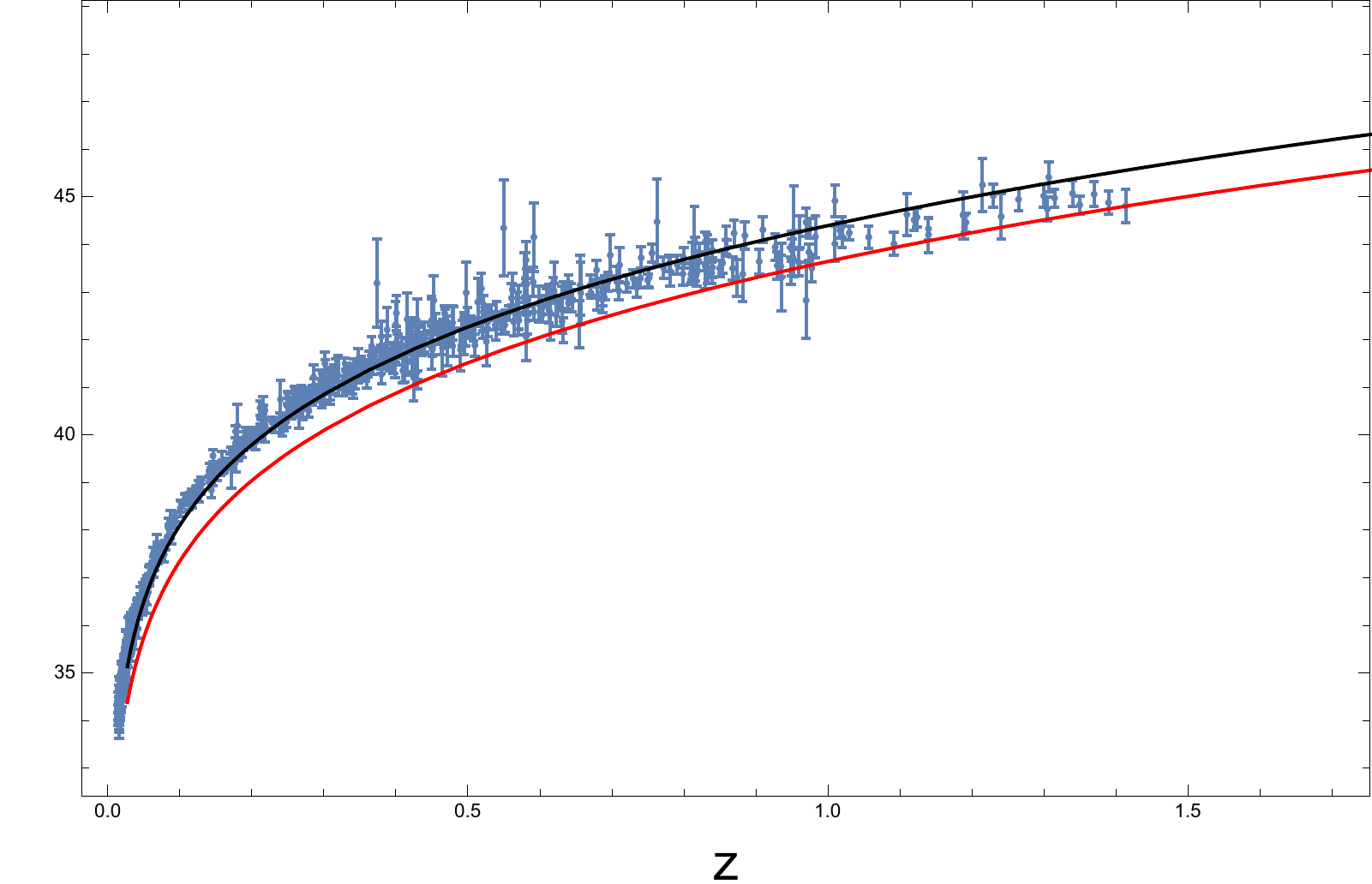}%
\caption{$\mu(z)$ vs $z$ plot for different values of $\delta$ along with observational data from Union 2 compilation, here $\delta=-0.0001$ (black) and $\delta=-0.0002$ (red).}%
\label{supernova}%
\end{figure}

\section{Discussion}
\label{sec:disc}

We have made a comparative study of RHDE models in different dimensions. A few points are worth noting. The density parameter is plotted in figs. (\ref{fig:param1}), fig. (\ref{fig:param2}), and (\ref{fig:param3}) for four, five and ten dimensions respectively taking three different delta values.  It is seen that density parameter exhibits a sharp transition towards a value close to unity as dimension is increased. For positive $z$, the variation in $\Omega_D$ with delta is seen from the figures, but in the present universe and in near future the behaviour becomes almost independent of model parameter $\delta$. This trend is same for all the dimensions considered.  Another thing to note here is that as the dimension is increased $\Omega_D$ becomes almost independent of $\delta$ throughout the evolution of the universe. From the plot of the eos parameter it is seen that dark energy begins in a quintessence era in the early universe and the value decreases reaching a value of $\omega_D=-1$, in the present universe. It remains constant at $\omega_D=-1$ for negative $z$ never crossing the phantom divide ($\omega_D<-1$). This behaviour is dimension independent.  We note that for $\delta=-0.0005$, the universe transitioned from a dust filled phase to quintessence era. The universe transits from a decelerated phase to accelerated phase as shown in figs. (\ref{fig:param1}), fig. (\ref{fig:param2}), and (\ref{fig:param3}). For $\delta=-0.0005$ the universe transits from decelerated to accelerated phase, however for higher delta universe remains in the ever-accelerating phase. As the dimension of the universe is increased higher $\delta$ values correspond to ever accelerating universe as is evident from the figures (1)-(3). For different value of the model parameter $\delta$, present value of $q$ is seen to vary from $-0.4$ to $-0.1$ which is very much acceptable in light of Sn-Ia data. Fig. (\ref{supernova}) depicts a plot of distance modulus ($\mu(z)$) with redshift for different values of the model parameter $\delta$ along with Union Compilation data \cite{union}. Clearly, some values of $\delta$ (e.g. $\delta = -0.0001$) is observationally favoured. A full scale analysis of observational constraints, however, is beyond the scope of the present work. The focus of this study has been the late time evolution of the universe only and the effect of higher dimension on this phase. However, it would be interesting to investigate some unified scenarios based on HDE \cite{nojiriunify,nojiriuorg} which also includes an early inflationary phase.


\end{document}